\documentclass[a4paper, 11pt]{article}
\usepackage{amsmath} \usepackage{amsfonts} \usepackage{amssymb}\usepackage{latexsym}
\usepackage[english]{babel}
\usepackage{amscd}
\usepackage[dvips,final]{graphics}
\usepackage{fancyhdr}
\usepackage{locengli,math}
\usepackage{graphicx,pst-plot,psfig,pstricks,pst-node,pst-text,pst-tree,pst-char,pst-coil}

\newcommand{\Ito}{\textrm{Ito}}

\begin{document}

\begin{center}
\textbf{\LARGE{\textsf{Commutativity and the third Reidemeister movement}}}
\footnote{
\textit{1991 Mathematics Subject Classification:}
\textit{Key words and phrases:}
}

\vskip1cm
\parbox[t]{14cm}{\large{
Philippe {\sc Leroux}}\\
\vskip4mm
{\footnotesize
\baselineskip=5mm
Institut de Recherche
Math\'ematique, Universit\'e de Rennes I and CNRS UMR 6625\\
Campus de Beaulieu, 35042 Rennes Cedex, France, philippe.leroux@univ-rennes1.fr}}
\end{center}

\vskip1cm
{\small
\vskip1cm
\baselineskip=5mm
\noindent
{\bf Abstract:}
In quantum information theory, for $a,b$ two positive operators living in
$B(\mathcal{H})$, where $\mathcal{H}$ is a separable Hilbert space, the quantum fidelity is denoted
by $a*b =(b^\frac{1}{2}ab^\frac{1}{2})^\frac{1}{2}$. One of the aim of this letter is to interpret the quantum fidelity
as an algebraic law. We remark that if $a,b,c$ are three positive operators which commute pairwise, the law $*$ becomes
self-distributive, i.e. the third Reidemeister movement in knot theory is verified. We study the converse. Let three positive
operators be given, does the fact that the third Reidemeister movement between them is possible implie that they commute pairwise ?
Though in general we only conjecture it for the moment, we prove it in some particular but important cases.
Should this movement be not possible, we interpret it as an obstruction to commutativity.
We give also new examples of quandle algebras and left distributive systems and study the generalisation of Ito maps.

\section{Introduction}
Quandle sets and left distributive products arise naturally in knot theory. Reidemeister showed that all the movements
we can do on a given knot can be decomposed into three
topological movements called the Reidemeister movements. Expressing them in an algebraic way, they yield the axioms for quandle
set and left distributive products. For the convenience of the reader, we give all the definitions and yield other examples of such sets.
Let $B(\mathcal{H})$ be the algebra of bounded operators acting on a separable Hilbert space. In quantum information theory, we can
define for positive operators, $a,b \in B(\mathcal{H}) $, the following product: $a*b =(b^\frac{1}{2}ab^\frac{1}{2})^\frac{1}{2}$.
If both of these operators are of trace one, they model quantum systems. In a family of mutually commuting positive operators,
we notice
that such a set, equipped with the quantum fidelity law, $*$, generates a self-distributive set, i.e. the third Reidemeister movement
is possible among the given positive operators. We conjecture that the reverse also holds, that is
the third Reidemeister movement
is possible between three given positive operators entails that these three operators commute pairwise. We prove this conjecture
for important cases and interpret the non possibility of such a movement as an obstruction to commutativity.
\section{Quandle algebras and LDRI systems}
\textbf{Notation}: All over this paper, we denote by $\mathbb{R}^*_+$, the set of strictly positive reals and by
$\mathbb{C}^*$, the set of complexes different from zero. The field $\mathbb{K}$ will denote either $\mathbb{R}$ or
$\mathbb{C}$.
\begin{defi}{[Quandle algebras and LDRI systems]}
Let $S$ be a set equipped with a product $*: S \times S \xrightarrow{} S$. $S$ is said a {\it{quandle algebra}} \cite{Fenn} if it verifies the axioms
$R_1, R_2, R_3$ or $R_1, R_2, R_{3'}$, with:
\begin{itemize}
\item[$R_1:$] {Idempotent law or the first Reidemeister movement, i.e. for all $a \in  S$, $a*a=a$.}
\item[$R_{2}:$]{The second Reidemeister movement, i.e for all $a,b \in S$, there exists a unique $c \in S$ such that $b=a*c$. }
\item[$R_{3}:$]{Left distributivity or the third Reidemeister movement, i.e for all $a,b,c \in S$, $a*(b*c) = (a*b)*(a*c)$.}
\item[$R_{3'}:$]{Right distributivity or the third Reidemeister movement, i.e for all $a,b,c \in S$, $(b*c)*a = (b*a)*(c*a)$.}
\end{itemize}
$S$ is said a {\it{left distributive (LD) system}} \cite{Deh} if it verifies at least the third axiom of a quandle algebra and a {\it{(LDI) system}} if it verifies
$R_1$ and $R_3$. A {\it{right distributive (RD) system}} $S$ verifies $R_{3'}$.
The definition of a {\it{LDRI system}} is now obvious.
\end{defi}
\Rk
We denote for all $a \in S$, $\Psi_a: S \xrightarrow{} S, \ x \mapsto   \Psi_a(x) = a*x$.
In terms of this function, for every $a \in S $, $R_3$ means that $\Psi_a$ is an $*$-homomorphism since for all $b,c \in S$ we get
$\Psi_a(b*c)=\Psi_a(b) * \Psi_a(c)$. $R_1$ means that any $\Psi_a$ has at least a fixed point $a$ and $R_{2}$ means that any
$\Psi_a$ is injective. Stated otherwise we can say that $*$ is left cancellative i.e. $a*c =a*c'$ implies $c=c'$.
\begin{defi}{[Entropic law law]}
In \cite{Deh} Dehornoy is interested in sets with particular products.
A set $S$ will be called an \textit{entropic} or a \textit{medial system} iff $S$ is equipped with a product such that all its elements verify $(xy)(uv)=(xu)(yv)$.
\end{defi}
\section{New examples}
\textbf{Convention}: Fix $z \in \mathbb{C^*}$ with argument $\theta \in [0, \ 2\pi[$ and $a \in ]0,1[$.
Among all the possible roots of $z^a$, we choose
$\vert z \vert^a \exp(i a \theta)$.

Few examples of LD systems are known in the literature. We present here some new examples.
\begin{exam}{[ $\mathbb{C}$ and $\mathbb{R_+}$]}
\begin{theo}
\label{cr}
Let $a,b \in \mathbb{R}$.
the map $\mathbb{C}\times \mathbb{C}\xrightarrow{\star} \mathbb{C}$,  $(y,z) \mapsto y^a z^b$ embeds $\mathbb{C}$
into a LDRI system and  $\mathbb{C}^*$ into a quandle algebra iff $a+b =1$. In this case $R_1 \Leftrightarrow R_{3}$.
If $a,b \in \mathbb{R_+}$, the same result holds, i.e.
$(\mathbb{R_+}, \star)$ is a LDRI set and $ (\mathbb{R^*_+}, \star)$ a quandle algebra. Moreover this product is entropic.
\end{theo}
\Proof
Straightforward.
\eproof
\Rk
In fact the group $\mathbb{C}^*$ is already known as a quandle algebra. Indeed equipped with the law $a \lhd b = a^{-1}b a^{-1}$,
any group can be embedded into a quandle algebra. We have just discovered another product which guaranties the same embbeding.
In some sense, we can say that $(\mathbb{C}^*, \star, \lhd)$ is a di-quandle algebra.

Before going on, we recall the Schur product \cite{Horn} of two matrices. Let $M_n(k)$ be the algebra of $n$ by $n$ matrices over the field $k$. Let
$a,b \in M_n(k)$, the Schur product of $a$ and $b$, denoted $a \circ b$ is defined by: $(a \circ b)_{i,j}= a_{i,j}b_{i,j}$. If $a$ and $b$ are
positive, so is $a \circ b$. In the sequel, $(a)^{\circ \frac{1}{2}}$ will mean the square root in the sense of the Schur product.
\begin{exam}{[Non negative matrices]}
Let $x \in M_n(\mathbb{R})$. The matrix $x$ is said non negative if all its components are non negative. We denote $NN$ the set of non negative
matrices.
Obviously we can embed $NN$ into a LDRI set by defining $x * y = x^{\circ a}y^{\circ b}$, with $a+b =1$ and $a,b$ positive.
\end{exam}
\begin{exam}{[Self-adjoint matrices]}
With the convention of the root of a complex number imposed at the begining of this section and by noticing that the Schur product
of two self-adjoint matrices is still self-adjoint, we can embedd the set of self-adjoint matrices into a LDRI set by the law
$x * y = x^{\circ a}y^{\circ b}$, with $a+b =1$ and $a,b$ positive.
\end{exam}
\section{A connection of Ito derivatives to the third Reidemeister movement: the distributivity defect of an Ito map}
In \cite{Coa}, we define a generalisation of an Ito derivative.
\begin{defi}{[Ito derivation]}
Let $A,B$ be two unital associative algebras.
A linear map $d_{\Ito} : A \xrightarrow{} B$ such that $\forall \ x,y \in A$:
\begin{enumerate}
\item {$d_{\Ito}(1_A) = 0_B,$}
\item {$d_{\Ito}(xy) =  d_{\Ito}(x)y + xd_{\Ito}(y) + d_{\Ito}(x)d_{\Ito}(y),$}
\end{enumerate}
is called an {\it{Ito derivative}}.
\end{defi}
We recall from \cite{Coa}, the following theorem.
\begin{theo}
Let $A$ be an unital algebra.
The set of Ito derivatives from $A$ to $A$ is in one to one with the set of homomorphisms from $A$ to $A$.
We have: $\rho + id = d_{\Ito}$ and $\rho - id = d_{\Ito}$, where $\rho$ and $d_{\Ito}$ are respectively a homomorphism and the associated
Ito map.
\end{theo}
In Quantum Mechanics, if $H \in B(\mathcal{H})$ is an hamiltonian, that is a self-adjoint operator, $H \mapsto [H, \cdot]$,
defines a Leibnitz derivative which is closely related to the evolution of any observable. Thus, in generalising this remark to any element of
$A$ we have for $a \in A$,  $[a, \cdot]: b \mapsto ab-ba$.
Hence,
$$L: A \xrightarrow{} \textsf{Hom}(A,A), \ \ \ a \mapsto [a, \cdot].$$
If we define $\circ$ such
that $a \circ b := [a, b]$ we get the Leibnitz identity:
$$ (x \circ y) \circ z - x \circ (y \circ z) = (x \circ z) \circ y ,$$
which means that we can control the lack of associativity of the product $\circ$. In the case
of Ito derivatives, can we embed $A$, into  $\textsf{Hom}(A,A)$, such that the maps
involved describe Ito derivative, instead of Leibnitz derivative ?

If we define $\star$ such that $x \star y := \rho_x(y)$, we get the Ito identity:
$$ x \star (y \cdot z) - (x \star y)\cdot(x \star z) = (x \star y)\cdot z  + y \cdot (x \star z),$$
which means that the lack of distributivity of the product $\star$ with regard to the
product $\cdot$ can be controlled.
This remark can be used to generalise the definition of the Ito derivative concept.
For example if $A$ is equipped with a product $\rhd$ which verifies the third Reidemeister movement,
i.e. $a \rhd (b \rhd c) = (a \rhd b) \rhd (a \rhd c)$ then for all elements of $A$, the map $
I: A \xrightarrow{} \textsf{Hom}(A,A)$ defined by,
$$ a \mapsto \rho_a(\cdot) = a \rhd (\cdot) - id := a \star (\cdot),$$
sends an element $a$ into an Ito derivative since $\Psi_a (c) := a \rhd c$ is a
$\rhd$-homomorphism and there is a bijection between Ito maps and homomorphisms. We will have:
$$ x \star (y \rhd z) - (x \star y) \rhd (x \star z) = (x \star y)\rhd z  + y \rhd (x \star z).$$

In some sense if $h$ is a homomorphism from $A$ to $A$ for the usual product, we can say that
$h$ verifies an (auto)-distributivity
condition since for all $a,b \in A$ we have, (if we define $h(a) := h \star a$),
$$ h \star (a \cdot b) - (h \star a)\cdot(h \star b) = h(ab) - h(a)h(b).$$
That is why there is a link between Ito derivatives and homomorphisms.

An example of such a situation is the complex field, $\mathbb{C}$ with the laws described in part three. Every point of $\mathbb{C}$
yields an Ito map.
\end{exam}
\section{Obstruction to commutativity}
This subsection is an attempt to generalize what was said in theorem \ref{cr} into a non commutative world. We will
see that the right generalisation of the $\star$ product is what is called the fidelity in quantum information theory. We will
show also that an obstruction of the third Reidemeister movement, for this new law, can be viewed as an obstruction to
the commutativity of positive operators.

Let $\mathcal{H}$ be a separable Hilbert space. In this paper, any operator considered will be bounded. We denote by $P(\mathcal{H})$ the set of bounded positive operators
and $P(\mathcal{H})_+$ the set of bounded strictly positive operators. Two brackets will be used, $[a,b] = ab-ba$ and $\{a,b\}=ab +ba$
for any $a,b \in \mathcal{L}(\mathcal{H})$. We denote by $ \{a \}' $ the commutant of $a$, i.e. the set of operators which commute
with $a$ and by $I$ the unit of $P(\mathcal{H})$. By $\mathbb{D}$ we mean the set of density operators, i.e. the
intersection of $P(\mathcal{H})$ with the set of trace one operators. By $\textsf{Spec}(a)$, we mean the spectrum of an operator $a$.
We denote by $a^\dagger$ the adjoint of the operator $a$.
\begin{defi}{[The $*$ product]}
Let $a,b \in P(\mathcal{H})$, we define the $*$ product by: $a*b =(b^\frac{1}{2}ab^\frac{1}{2})^\frac{1}{2}$.
\end{defi}
\Rk
Let $\lambda \in \mathcal{R}_+$, then $a*(\lambda b) = (\lambda a)*b = \sqrt{\lambda }(a*b)$.
\Rk
In quantum information theory, this product defines the fidelity of an operator $a$ with regard to $b$, when both $\tr \ a =1 =\tr \ b$.
In a Banach algebra, we know that $\textsf{Spec}(xy) \cup \{0\}=\textsf{Spec}(yx) \cup \{0\}$. Here
$\textsf{Spec}(a*b)(a*b)=\textsf{Spec}(b*a)(b*a)$, since if we denote $Y=a^\frac{1}{2}b^\frac{1}{2}$, we obtain $(a*b)(a*b)=YY^\dagger $ and
$(b*a)(b*a)=Y^\dagger Y$ and if $Y$ is not invertible, so is $Y^\dagger$. Since $t \in \mathcal{R_+} \mapsto \sqrt{t}$ is
a continous fonction on the spectra of $(a*b)(a*b)$ and $(b*a)(b*a)$, we get $\textsf{Spec}(a*b)=\textsf{Spec}(b*a)$. This implies that if $a,b$ are trace class operators, we get
$\tr(a*b)=\tr(b*a)$.
\begin{theo}
Suppose $P(\mathcal{H})$ is a commutative set, (classical case).
Then the product $*$  embeds $P(\mathcal{H})_+$ into a quandle algebra and $P(\mathcal{H})$ into a LDRI set.
\end{theo}
\Proof
As $P(\mathcal{H})$ is a commutative set, we have for all $a,b \in P(\mathcal{H})_+$,
$a*b =b*a=b^\frac{1}{2}a^\frac{1}{2}$. The rest is straightforward.
\eproof
\Rk
This theorem is important. It is the commutativity of the operators which garanties the braiding of them.
\begin{coro}
Let $a,b,c \in P(\mathcal{H})$. If the operators
$a,b,c$ commute pairwise then the third Reidemeister movement, between them, is possible.
\end{coro}
We study the converse of this theorem to prove that non commutativity of positive operators can be viewed as an obstruction to
the third Reidemeister movement.
In fact the converse is for the moment a conjecture. We prove in the sequel some particular, but important, cases. \\
\\
\textbf{Conjecture }   Let $(a,b,c) \in P(\mathcal{H})$.
If the third Reidemeister movement is possible between them,
whatever the position of $a,b,c$ in the equation of $R_{3}$, then $a,b,c$ have to commute pairwise. \\

For the study of some particular cases, we need the following theorem.
\begin{theo}
Let $X$ a normal operator. Then $X=BC$, where $B, \ C$ are self-adjoint operators, entails that
$B$ and $C$ commute.
\end{theo}
\Proof
Since $X$ is normal, it commutes with its adjoint. We have, say $Xx=\lambda_x x$, where $x \in \mathcal{H}$ and $\lambda_x$
a non null scalar.
Since $Xx \not=0$, $Cx \not=0$.
Moreover we notice that $CX = CBC$ is self-adjoint. Hence $ x^\dagger CX x =\lambda_x x^\dagger C x$. Since $C$ and $CX$
are self-adjoint operators,
all the eigenvalues of $X$ must be real. That is $X=X^\dagger$. In this case $BC=CB$.
\eproof

For $a,b \in P(\mathcal{H})$, we denote $[a,b]_*= a*b - b*a$.
\begin{theo}
Let $a,b \in P(\mathcal{H})$.
$$[a,b]_* = 0 \Leftrightarrow [a,b] = 0.$$
\end{theo}
\Proof
Let $a,b \in P(\mathcal{H})$.
The equation, $a*b=b*a$, is equivalent to (by unicity of the square root of a positive operator)  $b^\frac{1}{2}ab^\frac{1}{2}=a^\frac{1}{2}ba^\frac{1}{2}$,
i.e. $X :=a^\frac{1}{2}b^\frac{1}{2}$ is normal, i.e. $[a,b] = 0$.
\eproof
\begin{theo}
Let $a,b,c \in P(\mathcal{H})$, with $a,b$ strictly positive. If $[a,b] = 0$ and $[a,c] = 0$ then
the third Reidemeister movement between $(a,b,c)$ is possible iff $[b,c]=0$.
\end{theo}
\Proof
Among all the writtings of $R_{3}$, we choose to study what constraints imply the following choice:
$$ b*(a*c) = (b*a)*(b*c).$$
Since $[a,c]_*=0$, $[(b*a), (b*c)]_*=0$, i.e. $[(b*a), (b*c)]=0$, i.e. $[b, c]=0$.
\eproof
\Rk
This theorem claims that in the commutant of $a$, two operators are braidable, i.e. $R_{3}$ is possible, iff
$b$ and $c$ belong to $\{b \}^{' } \cap \{c\}^{'}$. Stated otherwise the non possibility of the third Reidemeister movement
can be viewed, in the commutant of $a$, as an obstruction to commutativity.
\begin{coro}
Let $b,c \in P(\mathcal{H})$, with $b$ invertible.
$$[b,c]=0 \Leftrightarrow (I,b,c) \ \ \textrm{are braidable.}$$
\end{coro}
\Rk
Since $[b,c]=0 \Leftrightarrow [b,c]_*=0$, then for all continuous functions $f$ on the compact \textsf{Spec}($b$) and $g$ on \textsf{Spec}($c$), we can claim
that if $(I,b,c)$ are braidable then so are $(I,f(b),g(c))$.
\begin{coro}
Let $b,c \in P(\mathcal{H})$, with $b$ invertible.
$$[b,c]=0 \Leftrightarrow b*c =b^\frac{1}{4}c^\frac{1}{2} b^\frac{1}{4}. $$
\end{coro}
\begin{theo}
Let $a,b,c \in P(\mathcal{H})$, with $a,b$ invertible. Suppose $[b,c] = 0$. Then
the third Reidemeister movement, between  $(a,b, b^\frac{1}{2}, c)$ is possible is equivalent to say that $(a,b,c)$ commute pairwise.
\end{theo}
\Proof
Possibility of $R_{3}$ between $(a,b,c)$ implies that, for instance $a*(b*c) = (a*b)*(a*c)$, i.e.
as $[b,c]_*=0$, $(a*b)*(a*c)=(a*c)*(a*b)$, i.e $ [a*b,a*c]_*=0$, i.e. $bac=cab$. Similarly,
$R_{3}$ is possible between $(a,b^\frac{1}{2},c)$ implies that $b^\frac{1}{2}ac=cab^\frac{1}{2}$, since $[b^\frac{1}{2},c] = 0$.
Hence $b^\frac{1}{2}(b^\frac{1}{2}ac)=(cab^\frac{1}{2})b^\frac{1}{2}$, i.e. $[a,c]=0$. Hence $[b,c] = 0$,  $[a,c]=0$ and as
$R_{3}$ is possible, this implies $[a,b] = 0$.
\eproof
\Rk
With the assumptions of this theorem, the impossibility of such a movement can be viewed as an obstruction to commutativity.

It is interesting to study all properties above on the density operators set $\mathbb{D}$. For convenience we include $0$ in $\mathbb{D}$.
\begin{defi}{}
Let $a,b \in \mathbb{D}$ such that $a*b \not= 0$. We denote $a \oslash b = \frac{a*b}{\tr (a*b)}$.
\end{defi}
\begin{theo}
Let $\mathcal{T}$, a set of mutually commuting positive operators. Then
$(\mathcal{T}, * )$ generates a LDRI system. If all the elements of $\mathcal{T}$ are invertible, then $(\mathcal{T}, *)$
generates a quandle algebra. Similarly, if  $\mathcal{T}_1 \subset \mathcal{D}$ is a set of mutually commuting density operators, then
$(\mathcal{T}, \oslash )$ generates a LDRI system. If all the elements of $\mathcal{T}$ are invertible, then $(\mathcal{T}, \oslash )$
generates a quandle algebra.
\end{theo}
\Proof
Straightforward for the first and third movement. For $R_{2}$, let $(a,b)$ be two positive operators. We
notice that there is a unique $c$ strictly positive such that $b = a*c$.
In the case of a family of density operators, i.e. if $a,b$ are density operators, we obtain that $c' :=\frac{c}{\tr(c)}$ is the unique density
operator such that $b = a \oslash c'$.
\eproof
\Rk
Let $a,b \in \mathcal{T}$, a set of mutually commuting strictly positive operators.
We have said that $a \lhd b = a^{-1}b a$ is also a  strictly positive operator and that the law $ \lhd $ embeds $\mathcal{T}$
into a quandle algebra.
\Rk \textbf{[Distributivity]}
Let $a,b,c \in \mathcal{T}$ be a set of mutually commuting positive operators. Then the operator product $ab$ is still positive.
We can as well study the set generated by $( \mathcal{T},*,m)$, where $m$ denotes the operator product. We have
$a(b*c) = (a*b)(a*c)$ and $(aI)*(bc)=a*(bc) = (a*I)(b*c)$. We notice also that $a*I=I*a=a^{\frac{1}{2}}$.
\begin{prop}
$a,b \in \mathcal{P}_+$.
Set $X^\frac{1}{2}=a^{-1}*(a*b)$ and $Y^\frac{1}{2}=b^{-1}*(b*a)$.
Then, $XY=YX=I$.
\end{prop}
\Proof
Recall that for any invertible bounded operator $Z$, the polar decomposition yields $Z =U(Z^{\dagger}Z)^\frac{1}{2}$, where
$U$ is a unitary operator. Similarly $UZ^{\dagger} =(ZZ^{\dagger})^\frac{1}{2}$. With $Z:=a^{\frac{1}{2}}b^{\frac{1}{2}}$.
$XY = a^{-\frac{1}{2}}(a*b)a^{-\frac{1}{2}}Y=a^{-\frac{1}{2}}(UZ^{\dagger})a^{-\frac{1}{2}}
b^{-\frac{1}{2}}(U^{\dagger}Z)b^{-\frac{1}{2}}=I$. Similarly for the other equality $YX=I$.
\eproof
\begin{theo} \textrm{[compatibility with the order structure in $\mathcal{P}$]}
Let $a,b,x \in \mathcal{P}$.
$$ a \leq b \Rightarrow x*a \leq x*b.$$
\end{theo}
\Proof
Let $a,b,x \in \mathcal{P}$. $0 \leq a \leq b \Rightarrow 0 \leq x^{\frac{1}{2}}a  x^{\frac{1}{2}} \leq x^{\frac{1}{2}}b x^{\frac{1}{2}}
\Rightarrow x*a \leq x*b.$
\eproof
\begin{theo}
Let $x,y,z \in \mathcal{P}$ be three mutually non orthogonal projectors of rank one.
The third Reidemeister movement is possible iff $x=y=z$.
Let $x_1,y_1,z_1 \in \mathcal{P}$ be three mutually non orthogonal (trace-class) operators of rank one, then the three operators, $x_1,y_1,z_1$ are proportional.
\end{theo}
\Proof
Let $e,f \in \mathcal{P}$ be two non orthogonal projectors of rank one. Then, $e^{\frac{1}{2}}=e$ and $efe =\tr(ef)f$.
This remark yields, for $x,y,z \in \mathcal{P}$, three mutually non orthogonal projectors of rank one, $\tr(xy)=\tr(xz)=\tr(yz)=1$.
However, the Schwartz inequality yields $\tr (ef) \leq ( \tr e)(\tr  f)$, with equality iff there exists $\lambda > 0, \ e = \lambda f$.
Since, $\tr \ x =\tr \ y=\tr \ z=1$, we get $x=y=z$. Now if  $e_1 \in \mathcal{P}$ is
an operator of rank one, then $e_1^2 = \tr (e_1)e_1$, hence the last assertion.
\eproof
\subsubsection{The Wigner-Yanase information}
\begin{defi}{[The Wigner-Yanase information]}
Let $\rho$ a density matrix, i.e. a positive operator of trace one. Wigner and Yanase \cite{WY} define a notion of (quantum)
{\it{information}} of a self-adjoint operator $k$ with regard to the quantum system $\rho$ by:
$$ S_{WY}(k \ \vert \ \rho) := \frac{1}{2}\tr[ \rho^\frac{1}{2},k][ \rho^\frac{1}{2},k].$$
\end{defi}
\Rk
Based on this idea, we define: $S_{WY}(k,l \ \vert \ \rho):= \frac{1}{2}tr[ \rho^\frac{1}{2},k][ \rho^\frac{1}{2},l]$.

In Quantum Mechanics, the information is obtained via the trace map. The aim of this map is to
authorise a commutativity at short distance, since $\tr (ab)= \tr (ba)$. It is important to check our algebraic relations in allowing
us this freedom.
\begin{defi}{}
Let $R$ be a set of operators equipped with a
product $\star$. This product is said {\it{quasi-distributive }} if it verifies:
$$ \forall a,b,c \in R, \ \  \tr (a \star (b \star c) - (a \star b) \star (a \star c)) =0.$$
\end{defi}
\begin{exam}{}
We define for all $a,b,c \in P(\mathcal{H})$, $a \star b := \frac{1}{2} \{a^\frac{1}{2},b^\frac{1}{2}\}.$
\end{exam}
\Rk
A priori, $a \star b$ is not positive, though hermitian. If $[a,b] =0$, then $a \star b = a*b$. The map $\star$ allows us to build
an algorithm. Let $a,b,c$, three positive operators, we define our algorithm by an obvious recurrence:
$a \star b := \frac{1}{2} \{a^\frac{1}{2},b^\frac{1}{2}\}$, $a \star (b \star c) := \frac{1}{2} \{a^\frac{1}{2}, \frac{1}{2} \{b^\frac{1}{4},c^\frac{1}{4}\}\}$, and so on.
\begin{theo}
The map $\star$ verifies $R_1$. Moreover, for all $a,b,c \in P(\mathcal{H})$,
$$\tr(a \star (b \star c) - (a \star b) \star (a \star c))
=  \frac{1}{4}tr[a^\frac{1}{4},b^\frac{1}{4}][a^\frac{1}{4},c^\frac{1}{4}]
= \frac{1}{2}S_{WY}(b^\frac{1}{4}, \ c^\frac{1}{4}\  \vert \ a^\frac{1}{2}).$$
\end{theo}
\Proof
Straightforward by tedious calculations.
\eproof
\Rk
The Wigner-Yanase metric measures the defect of $\star$-homomorphism at point $a$.
\section{The Bures distance}
In quantum information theory, several distances exist in the space of density operators, one of them is the Bures distance.
\begin{defi}{[The Bures distance]}
The {\it{Bures distance}} \cite{Bures}\cite{Nielsen}\cite{Ditt}  is the map:
$$\mathbb{D} \times \mathbb{D} \xrightarrow{} \mathbb{R}_+  \ \ \ \ \  (a,b) \mapsto d_B(a,b) := (2 - 2\tr (a*b))^\frac{1}{2}.$$
\end{defi}
In \cite{Rylov}, Rylov defines another concept to replace Riemannian geometry. The key idea is based on a concept used in
general relativity called the world function. This function can be obtained from metric space after removal of some constraints.
We call  $\sigma$-{\it{space}}, a set $V=(\sigma, \Omega)$, a set $\Omega$ equipped with a function
$\Omega \times \Omega \xrightarrow{\sigma}\mathbb{R}_+$, such that for all point $P,Q \in \Omega$, $\sigma(P,P)=0$ and
$\sigma(P, Q)=\sigma(Q, P)$.
For instance if $V=(\sigma, \Omega)$ is a metric space, with a metric $\rho$, for all point $P,Q \in \Omega$, define
$\sigma(P, Q)= \frac{1}{2} \rho^2(P,Q)$.
The idea behind what Rylov calls $\sigma$-space and T-geometry, is to reformulate all the theorems and concepts from Euclidean
or Riemannian geometry only in terms of the world function. Such descriptions will be called $\sigma$-immanent.

In our case, we notice that the Bures distance on density operators gives rise to a world
function $\mathbb{D} \times \mathbb{D} \xrightarrow{\sigma} [0,1]$ such that $\sigma_B(\rho_1, \rho_2) := 1 - tr(\rho_1* \rho_2)$.

Let $\mathcal{P}^n := {P_0, \ldots, P_n} \subset \mathbb{D}$ a  finite set. The basic elements of
T-geometry are finite $\sigma_B$-subspaces $M_n(\mathcal{P}^n)$ of the $\sigma$-space $(\sigma_B, \mathbb{D})$ . For example Rylov defines the squared
length $\vert M_n(\mathcal{P}^n) \vert^2$ as the real number $\vert M_n(\mathcal{P}^n) \vert^2 = F_n(\mathcal{P}^n)$,
where $F_n(\mathcal{P}^n)$ is the Gram's determinant for the $n$ vectors $P_0P_i$, $i \in (1, \ldots, n)$,
i.e.  = $F_n(\mathcal{P}^n) = \det \| (P_0P_i. P_0P_j) \|$,
with $(P_0P_i. P_0P_j) \equiv
\Gamma(P_0, P_i, P_j) \equiv \sigma_B(P_0,P_i) + \sigma_B(P_0,P_j) - \sigma_B(P_i,P_j), \forall \ i,j \in (1, \ldots n)$.

What we would establish, is a relation between commutativity of density operators, quandle 2-cocycle and
$\sigma_B$-othogonality.
We now introduce some definition from quandle (co)-homology developed by \cite{Carter} \footnote{see also the included references.}.

\begin{defi}{}
Let $X$ be a quandle. $C_n ^R(X)$ will denote the free abelian group generated by $n$-tuples $(x_1, \ldots, x_n)$ of elements of the quandle $X$.
Define a homomorphism $\partial_n: C_n ^R(X) \xrightarrow{} C_{n-1} ^R(X)$ by:
$$\partial_n(x_1, \ldots, x_n) =
\sum_{i=2}^{n} (-1)^i [(x_1, \ldots x_{i-1}, x_{i+1}, \ldots, x_n)
 - (x_1*x_i, \ldots x_{i-1}*x_i, x_{i+1}, \ldots, x_n) ], $$
for $n > 1$ and $\partial_n =0$ for $ n \leq 1$.
\end{defi}
By using the third Reidemeister movement, one proves that $C_*^R(X)= \{ C_n^R(X), \partial_n \}$ is a chain complex. In fact
if $X$ is just a LDI system, $C_*^R(X)$ is still a complex.
As $(\mathbb{R}, +)$ is an abelian group we can define what is the chain and cochainquandle
complexes $C_*^R(X; (\mathbb{R}, +)) = C_*^R(X) \otimes (\mathbb{R}, +)$, with boundary $\Delta := \partial \otimes id$ and
$C_R^*(X; (\mathbb{R}, +)) = \hom( C_*^R(X), (\mathbb{R}, +))$, with coboundary $\delta = \hom(\partial, (\mathbb{R}, +))$.

\begin{exam}{}
A quandle 2-cocycle $\phi$ satisfies, for a 3-chain $(x_1, x_2, x_3)$,
$(\delta_2(\phi))(x_1, x_2, x_3) = \phi(\partial_3(x_1, x_2, x_3)) =0$, i.e:
$$ \phi(x_1, x_3) + \phi(x_1*x_3, x_2*x_3) = \phi(x_1, x_2) + \phi(x_1*x_2, x_3).$$
\end{exam}
\Rk
A LDI system 2-cocycle can be defined with the same manner.

We define $\Xi := \mathbb{R} \times \ldots \times \mathbb{R} \xrightarrow{} \mathbb{R}$, $(r1, \ldots, r_n) \mapsto r_1 + \ldots + r_n$.
\begin{theo}
Let $(\mathbb{D}_0, *)$ be a set of commuting density operators, closed under the law $*$, i.e a LDRI system.
Define $ \phi := \Xi(tr, tr): C_2 \xrightarrow{} \mathbb{R}$. $\phi$ is a 2-cocycle iff all the
$\Gamma(\rho_1, \rho_2, \rho_3) =0$, with $\rho_i \in (\mathbb{D}_0, *)$ .
\end{theo}
\Proof
Straightforward.
\eproof

\noindent
\textbf{Acknowledgment:}
The author wishes to thank Dimitri Petritis for fruitful advice for the
redaction of this paper.

\bibliographystyle{plain}
\bibliography{These}

\end{document}